\documentclass[prb,twocolumn,showpacs]{revtex4}     
\usepackage{epsfig,amsmath}

\begin{document}

\title{Electric transport theory of Dirac fermions in graphene} 

\author{Xin-Zhong Yan,$^{1,2}$ Yousef Romiah,$^1$ and C. S. Ting$^1$}
\affiliation{$^{1}$Texas Center for Superconductivity, University of 
Houston, Houston, Texas 77204, USA\\
$^{2}$Institute of Physics, Chinese Academy of Sciences, P.O. Box 603, 
Beijing 100080, China}
 
\date{\today}
 
\begin{abstract}
Using the self-consistent Born approximation to the Dirac fermions under finite-range impurity scatterings, we show that the current-current correlation function is determined by four-coupled integral equations. This is very different from the case for impurities with short-range potentials. As a test of the present approach, we calculate the electric conductivity in graphene for charged impurities with screened Coulomb potentials. The obtained conductivity at zero temperature varies linearly with the carrier concentration, and the minimum conductivity at zero doping is larger than the existing theoretical predictions, but still smaller than that of the experimental measurement. The overall behavior of the conductivity obtained by the present calculation at room temperature is similar to that at zero temperature except the minimum conductivity is slightly larger. 
\end{abstract}


\maketitle

\section{Introduction}

Electronic transport properties of graphene have attracted much interest since the experimental measurements were performed recently.\cite{Novoselov,Geim,Zhang,Morozov} Many theoretical models for the electric transport in graphene were focused on the short-range impurity scatterings,\cite{Shon,Khveshchenko,McCann,Aleiner,Ziegler,Peres,Ostrovsky} but the predictions cannot describe the experimental observations that the electric conductivity of graphene linearly depends on the carrier concentration.\cite{Geim} For the charged impurity scatterings, some theoretical works including the numerical diagonalization of the finite-electron system \cite{Nomura} and the calculations using the Boltzmann formalism \cite{Hwang} have been performed. The obtained electric conductivity is in overall agreement with the experiment. These works show strong evidence that the charged impurities are responsible for the electronic transport properties in graphene. 

The Boltzmann transport theory for graphene is based on the one-band approximation,\cite{Hwang,MacDonald,Cheianov} which is different from the usual two-dimensional systems. Its validity may become questionable for Dirac fermions at small carrier concentrations and at finite temperatures. The graphene has a band structure analogous to the massless relativistic Dirac particle. At low carrier concentrations, the Fermi energy is close to the zero where the upper and lower bands touch each other. Particularly, at zero doping and finite temperature, we have particle and hole excitations in the upper and lower bands. In this case, charge carriers in both bands should contribute to the electric transport. Therefore, the development of a proper transport theory for the Dirac fermions is of fundamental importance. The method of using the current-current correlation function should be such a choice, but it has been applied only for short-range impurity scatterings. Because of the complex nature of the involving matrix algebra, this approach has not yet been extended to study the transport in graphene with impurities of finite-range potentials.    

\begin{figure} 
\centerline{\epsfig{file=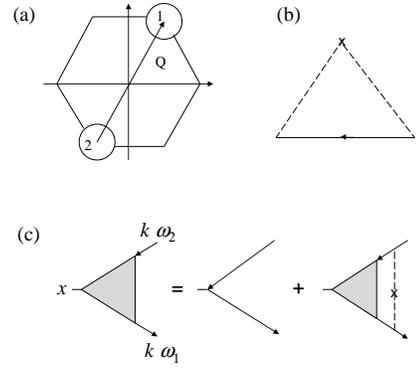,width=6.0 cm}}
\caption{(a) Brillouin zone and the two Dirac-cone valleys. (b) Self-consistent Born approximation for the self-energy. The solid line with arrow is the Green function. The dashed line is the effective impurity potential. (c) Current vertex with impurity insertions.}\label{fig1}
\end{figure} 

In this work, we present a new formalism for the electric transport of Dirac fermions under finite-range impurity scatterings based on the current-current correlation function. The current-vertex correction is shown to be determined by four-coupled integral equations. The two energy bands of the Dirac fermions are taken into account in this scheme. The present result should provide the more reasonable description of the electric transport of Dirac fermions at low doping and at finite temperature.

\section{Formalism}

We start with the Hamiltonian for electron-impurity interactions in graphene,
\begin{equation} 
H_1 = \sum_{j}\int d\vec Rn(\vec r_j)v_i(|\vec
 r_j-\vec R|)n_i(\vec R) ,
\end{equation}
where $n(\vec r_j)$ is the density operator of electrons at site-$j$ of the honeycomb lattice, $n_i(\vec R)$ is the real space density distribution of impurities, and $v_i(|\vec r_j-\vec R|)$ is the impurity scattering potential. For the situations related to low energy levels, electrons can be described by the Dirac fermions. The energy bands are given by two Dirac cones at the corners of the hexagon Brillouin zone. By noting this fact, we separate $H_1$ in momentum space into two parts: intravalley scatterings (within the same Dirac cone) and the intervalley ones (between the different Dirac cones). Using the Pauli matrices $\sigma$'s and $\tau$'s to coordinate the electrons in the two sublattices and two valleys, and suppressing the spin subscripts for briefness, the total Hamiltonian is given by
\begin{equation}
H = \sum_{k}\psi^{\dagger}_{k}v\vec
 k\cdot\vec\sigma\tau_3\psi_{k}+\frac{1}{V}\sum_{kq}\psi^{\dagger}_{k-q}V_i(q)\psi_{k} \label{H}
\end{equation}
where $\psi^{\dagger}_{k}=(c^{\dagger}_{ka1},c^{\dagger}_{kb1},c^{\dagger}_{ka2},c^{\dagger}_{kb2})$ is the electron operator with $a$ and $b$ denoting the sublattice and 1 and 2 for the valley indices, the momentum $k$ is measured from the center of each valley, $v$ ($\sim$ 5.856 eV\AA) is the Fermi velocity of electrons, $V$ is the two-dimensional volume of system, $V_i(q)=\sum_n \phi(q+ Q_n)$ with $Q_n$ the reciprocal honeycomb-lattice vector (where the summation over $Q_n$ is the result of separating the Fourier integral of the impurity potential over the whole momentum space into Brillouin zones), and $\phi(q)$ is given by 
\begin{equation}
\phi(q) = 
\begin{pmatrix}
n_i(-q)v_i(q)\sigma_0& n_i(Q-q)v_i(q-Q)\sigma_0 \\
n_i(-Q-q)v_i(q+Q)\sigma_0& n_i(-q)v_i(q)\sigma_0 
\end{pmatrix} \nonumber
\end{equation}
with $Q$ a vector from the center of valley 2 to that of the valley 1, and $\sigma_0$ is the $2\times 2$ unit matrix. Here, all the momenta are understood as vectors. A sketch of the Brillouin zone and valleys is shown in Fig. \ref{fig1}(a). From our previous result,\cite{Yan} the cutoff of $k$ for $k$-summation is about $k_c \sim \pi/3$ (in unit of the lattice constant $a$ = 1) within which the electrons can be regarded as Dirac particles. The momentum transfer $q$ is constrained so that an electron at $k$ is scattered to $k+q$ ($k+q+Q$) in the same (different) valley. Within the validity of the Dirac-fermions description for graphene, the carrier concentrations should be low and the radius of the Fermi circle is thereby small. Since the most important momentum transfer is about the order of the diameter of Fermi circle, for low energy excitations, $q$ is small. Therefore, the off-diagonal elements $v_i(q+ Q +Q_n)$'s can be considered as constants independent of $q$. Similarly, for the diagonal part, we have $v_i(q+Q_n)\approx v_i(Q_n)$ for $Q_n\ne 0$. Within the self-consistent Born approximation, after the average over the random impurity distributions, the impurity potentials will appear in the final result as 
\begin{equation}
\frac{1}{V}\sum_{n}\langle n_i(q+Q_n)n_i(-q-Q_n)\rangle
 v^2_i(q+Q_n)=n_i\sum_{n}v^2_i(q+Q_n)\nonumber
\end{equation}
where $n_i$ is the average impurity density. We can then define the effective potentials $v_0(q)$ and $v_1$ for the intravalley and intervalley scatterings respectively by 
\begin{eqnarray}
v^2_0(q)&=&\sum_{n}v^2_i(q+Q_n)\approx v^2_i(q)+\sum_{n\ne 0}v^2_i(Q_n), \nonumber\\
v^2_1 &=&\sum_{n}v^2_i(Q_n-Q). \nonumber
\end{eqnarray}
With these effective potentials, one needs to consider only one component without the summation over all $Q_n$. 

To analyze the electric transport, we firstly evaluate the Green function. We here use the self-consistent Born approximation (SCBA) \cite{Gorkov,Fradkin,Lee} that is shown in Fig. \ref{fig1} (b). Since the effective potentials are isotropic functions of $q$, the self-energy $\Sigma(\vec k,\omega)$ can be expressed as $\Sigma_0(k,\omega)\tau_0\sigma_0+\Sigma_c(k,\omega)\tau_3\hat k\cdot\vec\sigma$ with $\hat k$ the unit vector in $\vec k$ direction. We will occasionally drop the unity matrix $\tau_0\sigma_0$ for briefness. The Green function $G(\vec k,\omega)$ and the self-energy $\Sigma(\vec k,\omega)$ are determined by
\begin{eqnarray}
G(\vec k,\omega) &=& \frac{\tilde \omega + h_k\tau_3\vec\sigma\cdot\hat
 k}{\tilde\omega^2-h_k^2}   \label{sc1}\\
\Sigma_0(k,\omega) &=& \frac{n_i}{V}\sum_{k'}[v^2_0(|\vec k-\vec
 k'|)+v^2_1]\frac{\tilde\omega}{\tilde\omega^2-h_{k'}^2}\label{sc2}\\ 
\Sigma_c(k,\omega) &=& \frac{n_i}{V}\sum_{k'}v^2_0(|\vec k-\vec
 k'|)
\frac{h_{k'}\hat k\cdot\hat k'}{\tilde\omega^2-h_{k'}^2}\label{sc3}
\end{eqnarray}
where $\tilde\omega=\omega+\mu-\Sigma_0(k,\omega)$ with $\mu$ the chemical potential, $h_k = vk+\Sigma_c(k,\omega)$, and the frequency $\omega$ is understood as a complex quantity with infinitesimal small imaginary part. 

The current operator is $v\tau_3\vec\sigma$. The $x$-direction current vertex $v\Gamma_x(\vec k,\omega_1,\omega_2)$ satisfies the following $4\times 4$ matrix equation,
\begin{widetext}
\begin{equation}
\Gamma_x(\vec k,\omega_1,\omega_2)=
 \tau_3\sigma_x+\frac{1}{V^2}\sum_{k'}\langle V_i(\vec k-\vec k')G(\vec k',\omega_1)\Gamma_x(\vec
 k',\omega_1,\omega_2)G(\vec k',\omega_2)V_i(\vec k'-\vec k)\rangle,\label{vt}
\end{equation}
\end{widetext}
where $\langle\cdots\rangle$ means the average over the impurity distributions. This equation is shown diagrammatically in Fig. \ref{fig1}(c). It satisfies the Ward identity under the SCBA. To solve this equation, we analyze the structure of $\Gamma_x$. Firstly, since the outgoing and incoming momenta are the same $\vec k$ belonging to the same valley, the vertex matrix $\Gamma_x(\vec k,\omega_1,\omega_2)$ is diagonal in the valley space. In the right hand side of Eq. (\ref{vt}), except for $V_i$, all others are diagonal in the valley space. Because of $\langle n_i(q+Q)n_i(-q'-Q)\rangle/V = n_i\delta_{qq'}$, the off-diagonal elements of $V_i$ always appear in the right hand side of Eq. (\ref{vt}) as pairs: $\langle V_{i,\mu\nu}(\vec k-\vec k')V_{i,\nu\mu}(\vec k'-\vec k)\rangle$ (with $V_{i,\mu\nu}$ and $V_{i,\nu\mu}$ as respectively the $\mu\nu$th and $\nu\mu$th elements of $V_i$). Even $V_i$ is not diagonal, the average over the impurity distributions leads to the diagonal form in the valley space. Therefore, the diagonal form of $\Gamma_x$ is not changed by the impurity insertions. Secondly, supposing Eq. (\ref{vt}) is solved by iteration, one finds that only the matrices 
\begin{eqnarray}
A^x_0(\hat k) &=& \tau_3\sigma_x,\nonumber\\
A^x_1(\hat k) &=& \sigma_x\vec\sigma\cdot\hat k, \nonumber\\
A^x_2(\hat k) &=& \vec\sigma\cdot\hat k\sigma_x,\nonumber\\
A^x_3(\hat k) &=& \tau_3\vec\sigma\cdot\hat k\sigma_x\vec\sigma\cdot\hat k \nonumber 
\end{eqnarray}
are involved in the operations. For example, the result of the first-round iteration contains only these matrices. No other matrices can be generated in the further interactions. That is to say these matrices form a complete basis for the vertex $\Gamma_x$. To see this, we need prove that the resulted matrix of the multiplication of any one of these matrices by $\tau_3\vec\sigma\cdot\hat k$ [which appears in $G(\vec k,\omega)$] from both sides belongs to the same assemble. Actually, the matrix multiplications are given by
\begin{eqnarray}
\tau_3\vec\sigma\cdot\hat k(A^x_0,A^x_1,A^x_2,A^x_3) = (A^x_2,A^x_3,A^x_0,A^x_1),\\
(A^x_0,A^x_1,A^x_2,A^x_3)\tau_3\vec\sigma\cdot\hat k = (A^x_1,A^x_0,A^x_3,A^x_2).
\end{eqnarray}
Therefore, we can expand the vertex function as 
\begin{equation}
\Gamma_x(\vec k,\omega_1,\omega_2)=\sum_jy_j(k,\omega_1,\omega_2)A^x_j(\hat k), \label{vtc}
\end{equation}
which means 
\begin{equation}
y_j(k,\omega_1,\omega_2)=\int_0^{2\pi} d\phi{\rm Tr}[A^{x\dagger}_j(\hat k)\Gamma_x(\vec k,\omega_1,\omega_2)]/8\pi, 
\end{equation}
with $\phi$ the angle of $\vec k$. From Eq. (\ref{vt}), we obtain the equations determining the coefficients $y_j$'s,  
\begin{widetext}
\begin{equation}
y_j(k,\omega_1,\omega_2)= \delta_{j0}+\frac{1}{V}\sum_{k'j'}U_j(|\vec
 k-\vec k'|)L_{jj'}(k',\omega_1,\omega_2)y_{j'}(k',\omega_1,\omega_2)
  \label{yj}
\end{equation}
where
$U_0(q)=n_i[v^2_0(q)-v^2_1]$,
$U_1(q)=U_2(q)=n_iv^2_0(q)\cos\theta$,
$U_3(q)=n_iv^2_0(q)\cos 2\theta$ with $q = |\vec
 k-\vec k'|$, $\theta$ is the angle between $\vec k$ and $\vec k'$, and
\begin{equation}
L_{jj'}(k,\omega_1,\omega_2)=\int_0^{2\pi}\frac{d\phi}{8\pi}{\rm Tr}[A^{x\dagger}_j(\hat k)G(\vec
 k,\omega_1)A^x_{j'}(\hat k)G(\vec k,\omega_2)].
\end{equation}
\end{widetext}
By expressing the Green function in the form
\begin{equation}
G(\vec k,\omega_l) = g_{0l} + g_{cl}\tau_3\vec\sigma\cdot\hat k,
\end{equation}
the matrix $L$ can be obtained as
\begin{equation}
L = 
\begin{pmatrix}
  g_{01}g_{02} &g_{01}g_{c2} &g_{c1}g_{02} &g_{c1}g_{c2}\\
	g_{01}g_{c2} &g_{01}g_{02} &g_{c1}g_{c2} &g_{c1}g_{02}\\
  g_{c1}g_{02} &g_{c1}g_{c2} &g_{01}g_{02} &g_{01}g_{c2}\\
	g_{c1}g_{c2} &g_{c1}g_{02} &g_{01}g_{c2} &g_{01}g_{02}\\	
\end{pmatrix}.\nonumber
\end{equation}

According to the Kubo formalism, the imaginary-time current-current correlation function $\Pi_{\mu\nu}(\tau)$ is defined as
\begin{eqnarray}
\Pi_{\mu\nu} (\tau) &=& -\frac{2}{V}\langle T_{\tau}J_{\mu}(\tau)J^{\dagger}_{\nu}(0)\rangle 
\end{eqnarray}
where $J_{\mu}(\tau) = -\sum_{\vec k}\psi^{\dagger}_{\vec k}(\tau)v\tau_3\sigma_{\mu}\psi_{\vec k}(\tau)$ is the $\mu$the component of the current per spin, and the factor 2 takes care of the spin freedom. Using the definition of $J_{\mu}(\tau)$, we express $\Pi_{\mu\nu} (\tau)$ as
\begin{eqnarray}
\Pi_{\mu\nu} (\tau) = \frac{2v^2}{V}\sum_{\vec k\vec k'}{\rm Tr}\langle T_{\tau}\psi_{\vec k'}(0)\psi^{\dagger}_{\vec k}(\tau)\tau_3\sigma_{\mu}\psi_{\vec k}(\tau)\psi^{\dagger}_{\vec k'}(0)\tau_3\sigma_{\nu}\rangle. \nonumber
\end{eqnarray}
To the lowest order in $n_i$, in the frequency space, $\Pi_{\mu\nu}$ is given by
\begin{widetext}
\begin{eqnarray}
\Pi^0_{\mu\nu} (i\Omega_m) = \frac{2v^2T}{V}\sum_{\vec k,n}{\rm Tr}G(\vec k,i\omega_n)\tau_3\sigma_{\mu}G(\vec k,i\omega_n+i\Omega_m)\tau_3\sigma_{\nu} 
\end{eqnarray}
\end{widetext}
where $T$ is the temperature, and $\omega_n$ and $\Omega_m$ are the fermion and boson Batsubara frequencies, respectively. With the impurity insertions under the conserving approximation consistent with the SCBA to the single particle Green function, $\Pi_{\mu\nu} (i\Omega_m)$ is obtained as
\begin{widetext}
\begin{eqnarray}
\Pi_{\mu\nu} (i\Omega_m) &=& \frac{2v^2T}{V}\sum_{\vec k,n}{\rm Tr}G(\vec k,i\omega_n)\Gamma_{\mu}(\vec k,i\omega_n,i\omega_n+i\Omega_m)G(\vec k,i\omega_n+i\Omega_m)\tau_3\sigma_{\nu} \nonumber\\
&\equiv & T\sum_n P_{\mu\nu} (i\omega_n,i\omega_n+i\Omega_m).   
\end{eqnarray}
\end{widetext}

The conductivity $\sigma$ is given by \cite{Mahan}
\begin{equation}
\sigma =
 \int_{-\infty}^{\infty}\frac{d\omega}{2\pi}[-\frac{dF(\omega)}{d\omega}][P(\omega^-,\omega^+)-{\rm Re} P(\omega^+,\omega^+)]\nonumber
\end{equation}
where $F(\omega)$ is the Fermi function, and $P(\omega_1,\omega_2) \equiv P_{xx}(\omega_1,\omega_2)$ is obtained as
\begin{eqnarray}
P(\omega_1,\omega_2)=
 \frac{2v^2}{V}\sum_{k}{\rm Tr}[A^x_0G(\vec k,\omega_1)\Gamma_x(\vec k,\omega_1,\omega_2)G(\vec k,\omega_2)] \nonumber 
\end{eqnarray}
and $\omega^{\pm} = \omega\pm i0$. Using Eq. (\ref{vtc}), we get
\begin{eqnarray}
P(\omega_1,\omega_2)=
 \frac{8v^2}{V}\sum_{kj}L_{01}(k,\omega_1,\omega_2)y_j(k,\omega_1,\omega_2). \nonumber 
\end{eqnarray}
For $\omega_1 = \omega^-$, and $\omega_2 = \omega^+$, $y_0$ and $y_3$ are real, $y_1 = y_2^{\ast}$, and $P(\omega^-,\omega^+)$ can be shown to be real. On the other hand, by using the Ward identity
\begin{equation}
\Gamma_x(\vec k,\omega^+,\omega^+)=
 \tau_3\sigma_x+\frac{\partial}{v\partial k_x}\Sigma(\vec k,\omega^+),
\end{equation}
the function $P(\omega^+,\omega^+)$ can be obtained explicitly
\begin{equation}
P(\omega^+,\omega^+)= \frac{1}{2\pi}{\rm Tr}[v\vec                            
 k_c\cdot\vec\sigma\tau_3 G(\vec k_c,\omega^+)].
\end{equation}
For the case of $\mu \ll vk_c$ and the magnitude of the self-energy $\ll vk_c$, the term $-{\rm Re}P(0^+,0^+)/2\pi$ contributes a value $\sim 2/\pi$ (in unit of $e^2/h$) independent of the doping to the zero-temperature conductivity. This is part of the minimum conductivity at zero doping, but it is missing in the one-band Boltzmann theory. Since the vertex correction is now determined by the four-coupled integral equations (6), the upper and lower energy bands of the Dirac fermions are automatically taken into account by the Green function. 

The Boltzmann formalism corresponds to the one-band approximation without intervalley scatterings ($v_1 = 0$). For electron doping, the conduction band is the upper band. By the upper band approximation, the Green function reads \begin{equation}
G(\vec k,\omega)\approx (1+\hat
 k\cdot\vec\sigma\tau_3)/2(\tilde\omega-h_k), 
\end{equation}
and the function $L_{jj'}(k,\omega_1,\omega_2)$ reduces to 
\begin{equation}
L_{jj'}(k,\omega_1,\omega_2) \approx G_+(k,\omega_1)G_+(k,\omega_2)/4 
\end{equation}
with $G_+(k,\omega) = 1/[\omega+\mu-vk-\Sigma_+(k,\omega)]$. The self-energy $\Sigma_+(k,\omega)\equiv\Sigma_0(k,\omega)+\Sigma_c(k,\omega)$ is determined by 
\begin{equation}
\Sigma_+(k,\omega)=
 \frac{n_i}{2V}\sum_{k'}v^2_0(|\vec k-\vec k'|)(1+\cos\theta) G_+(k',\omega).  \nonumber
\end{equation}
The vertex function $\Gamma_x(\vec k,\omega_1,\omega_2)$ is now related to only one function $z(k,\omega_1,\omega_2) =  \sum_jy_j(k,\omega_1,\omega_2)$. The latter is determined by the equation obtained by summation of Eq. (\ref{yj}) over $j$:
\begin{widetext}
\begin{equation}
z(k,\omega_1,\omega_2)= 1+\frac{n_i}{2V}\sum_{k'}v^2_0(|\vec k-\vec
 k'|)(1+\cos\theta)\cos\theta
 G_+(k',\omega_1)G_+(k',\omega_2)z_(k',\omega_1,\omega_2).  \label{z}
\end{equation}
\end{widetext}
By the further approximation $G_+(k',0^-)G_+(k',0^+)\approx -\pi\delta(E_F-vk){\rm Im}\Sigma_+(k,0^+)$ with $E_F = vk_F$ as the Fermi energy, one can obtain exactly the zero-temperature Boltzmann result.\cite{MacDonald,Cheianov,Hwang,Mahan}

Another case is the artificial point-contact impurity model. In this model, $v_0(q)=v_0$ is a constant, and $v_1 = 0$. Now in Eq. (\ref{yj}), except for $U_0$, all other angle integrals of $U_j$ vanish. Therefore, $y_0$ is the only relevant function in question. The function turns to be independent of the momentum and can be solved as
$y_0(\omega_1,\omega_2) = [1-U_0c(\omega_1,\omega_2)]^{-1}$
with 
\begin{equation}
c(\omega_1,\omega_2) = \frac{1}{V}\sum_{k}L_{00}(k,\omega_1,\omega_2).
\end{equation}
The function $P(\omega_1,\omega_2)$ is obtained as
\begin{equation}
P(\omega_1,\omega_2) =
 \frac{8v^2c(\omega_1,\omega_2)}{1-U_0c(\omega_1,\omega_2)},
\end{equation}
which coincides with the existing result.\cite{Ostrovsky} For the real zero-range impurity scatterings, $v_0 = v_1$, $U_0 = 0$ implying no vertex correction from the impurity insertions, one obtains $P(\omega_1,\omega_2) =
8v^2c(\omega_1,\omega_2)$. 

\begin{figure} 
\centerline{\epsfig{file=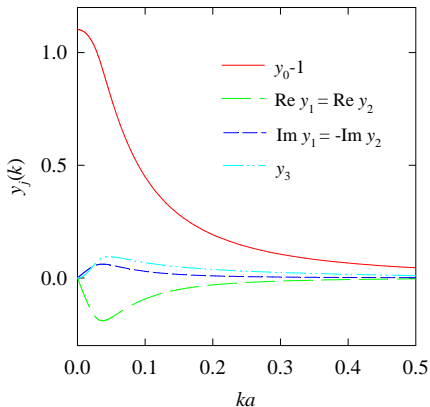,width=6.0 cm}}
\caption{(Color online) Functions $y_j(k) \equiv y_j(k,0^-,0^+)$ for $T = 0$ and $\delta = 1.0\times 10^{-4}$.}\label{cnd}
\end{figure} 

\section{Result}

The experimental observations of the electric transport in graphene have been previously analyzed.\cite{Peres,Hwang,Nomura} It is indicated that the charged impurities are the predominant scatters in graphene. To test our theory, we calculate the electric conductivity in graphene and compare it with the experimental results. In our numerical calculation, we adopt the charged impurity potential as the Thomas-Fermi type, $v_i(q)=2\pi e^2/(q+q_{TF})\epsilon$ where $q_{TF}=2\pi e^2\chi/\epsilon$, $\epsilon\sim 3$ is the dielectric constant due to the substrate electrons screening, and $\chi$ is the long-wave-length-limit static polarizability of the non-interacting electron system defined by
\begin{eqnarray}
\chi = \frac{2}{V}\sum_{\vec k}\int_0^{\beta}d\tau\langle T_{\tau}n(\tau)n^{\dagger}(0)\rangle_c
\end{eqnarray}
with $n(\tau) = \sum_{\vec k}\psi^{\dagger}_{\vec k}(\tau)\psi_{\vec k}(\tau)$. Here $\langle\cdots\rangle_c$ means that all the Green functions in the Feynman diagram are connected, and the factor 2 again comes from the spin freedom. By using the Green function of the non-interacting Dirac fermions, $\chi$ at low temperature $T$ is calculated as
\begin{eqnarray}
\chi &=& -\frac{2T}{V}\sum_{\vec k,n}{\rm Tr}G^0(\vec k,i\omega_n)G^0(\vec k,i\omega_n)\nonumber\\
&=& \frac{2\mu}{\pi v^2}[1+\frac{2T}{\mu}\ln(1+e^{-\mu/T})].
\end{eqnarray}
 The chemical potential $\mu$ is determined by
\begin{equation}
\frac{S}{V}\sum_{k}[F(vk-\mu)+F(-vk-\mu)-1] = \delta, \label{chm}
\end{equation}
where $S = \sqrt{3}a^2/2$ is the unit-cell area of the honeycomb lattice with $a \sim$ 2.4 \AA~ the lattice constant, and $\delta$ is the doped electron concentration per site. At $T = 0$, $\chi = 2k_F/\pi v$, and the Fermi wavenumber $k_F$ is determined by  $k^2_F = 4\pi\delta/\sqrt{3}a^2$. For low carrier doping concentrations, $q_{TF}$ is small and the effective potential $v_0(q)$ comes mainly from its leading term, $v_0(q)\approx v_i(q)$. For the off-diagonal part $v_1$, we use simply its leading order $v_1 \approx v_i(\overline{Q})$ with $\overline{Q} = 4\pi/3a$. The impurity density is chosen as $n_i = 1.15\times 10^{-3}a^{-2}$. 

Before showing the conductivity, we firstly present the numerical results of $y_j(k,0^-,0^+)\equiv y_j(k)$ the solution to Eq. (\ref{yj}) in Fig. 2 for $T = 0$ and $\delta = 1.0\times 10^{-4}$. These functions reveal how the current vertex is renormalized. By comparing to the bare vertex for which only $y_0(k) = 1$ is finite, it is seen that besides the $A^x_0$ component of the vertex is largely enhanced, the other $A^x_j$ ($j \ne 0$) components are generated by the impurity scatterings. These functions have appreciable magnitudes around the Fermi wavenumber.

Shown in Fig. \ref{cnd} is the comparison of the calculated electric conductivity with the experimental data.\cite{Geim} The conductivity $\sigma$ obtained by the theoretical calculation for both zero and room temperature linearly depends on $\delta$. This feature is in overall agreement with the experiment. With increasing $T$, $\sigma$ is enhanced at small doping while it is lowered at large doping. This is because as a function of $T$ the polarizability $\chi$ has a minimum around the chemical potential, which means that the screening effect is increased with increasing $T$ at small doping and the opposite at large doping. Especially, at zero doping $\mu = 0$ (due to the particle-hole symmetry), we have $\chi = 4(\ln 2) T/\pi v^2$; the finite $\chi$ comes from the particle-hole excitations. Due to the screening effect from the finite-temperature particle-hole excitations, the minimum conductivity $\sigma_{\rm min}$ at zero doping is increased from approximately 1.7 $e^2/h$ at $T = 0$ to $2.1 e^2/h$ at $T = 300$ K (both of them obtained by extrapolation from the results of finite carrier concentrations). Close to zero doping, the theoretical calculation gives rise to a smooth curve due to the appreciable inter-band mixing effect, especially at finite temperature. This feature is in agreement with the experimental observation that $\sigma$ is saturated at $\delta \to 0$. At very low carrier concentrations, the difference between the present calculation and the experiment is seen in the inset in Fig. \ref{cnd}. (For comparison, we note that the minimum conductivity predicted by the zero-range impurity scattering model is $4/\pi \approx 1.27$ in unit of $e^2/h$.\cite{Ostrovsky1}) We argue that the comparison with experiments could be improved if the effect of Coulomb interactions between electrons is considered. In our recent work based on the renormalized-ring diagram approach,\cite{Yan} considerable number of particle and hole excitations is shown to exist respectively in the upper and lower bands even at zero doping. The presence of these excited charge carriers not only imply the finite carrier density, but also give rise to effective screenings to the charged impurities and thus enhance the magnitude of the minimum conductivity as compared to what obtained in the present work. However, the incorporation of such an idea into the current-current correlation function is a difficulty task, and could be a subject for future study. Other possible explanation to the experimental result has been given by Hwang {\it et al.}\cite{Hwang} based on the inhomogeneity of the impurity distributions, and the existence of large carrier density fluctuations in the system. 

\begin{figure} 
\centerline{\epsfig{file=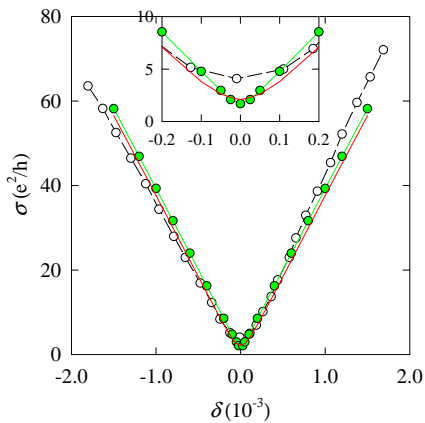,width=6.0 cm}}
\caption{(Color online) Electric conductivity $\sigma$ as function of electron doping concentration $\delta$. The present results (green circles for $T = 0$ and red solid line for $T = 300$ K) are compared with the experimental data (white circles). The inset is a magnification of the graph around zero doping.}\label{cnd}
\end{figure} 

\section{Summary}

In summary, we have presented the transport theory of Dirac fermions in graphene. For the first time, the current-current correlation function under impurity scatterings with finite-range potentials has been studied in the self-consistent Born approximation. The electric transport is described by four-coupled integral equations. The contributions of the charge carriers from both the upper and lower bands are included, which is essential for studying the transport properties of a Dirac-fermion system with low doping and at finite temperature. As a test of the present approach, we calculate the conductivity for graphene with charged impurities at zero and room temperatures. The obtained results are qualitatively consistent with experiments \cite{Geim} and the numerical diagonalization of finite size systems.\cite{Nomura} 

\acknowledgments

This work was supported by a grant from the Robert A. Welch Foundation under No. E-1146, the TCSUH, the National Basic Research 973 Program of China under grant No. 2005CB623602, and NSFC under grant No. 10774171.

\end{document}